\documentclass[aps,prb,twocolumn,superscriptaddress,showpacs,amsmath,amssymb,longbibliography]{revtex4-2}
\usepackage[english]{babel}
\usepackage{amsmath,amssymb,bm}
\usepackage{ulem}
\usepackage[nice]{nicefrac}
\usepackage{bm}
\usepackage{graphicx,bbm} 
\usepackage{times}
\usepackage{epsfig}
\usepackage[colorlinks,linkcolor=blue,citecolor=blue,urlcolor=blue]{hyperref}
\usepackage{color}
\definecolor{nred} {RGB}{224,0,0}
\definecolor{nblue} {RGB}{28,130,185}
\definecolor{dgreen} {RGB}{78,138,21}
\definecolor{norange}{RGB}{230,120,20}

\begin{document} 
\title{Modelling sample-to-sample fluctuations of the gap ratio in finite disordered spin chains}
\author{Bartosz Krajewski}
\affiliation{Department of Theoretical Physics, Faculty of Fundamental Problems of Technology, Wroc\l aw University of Science and Technology, 50-370 Wroc\l aw, Poland}
\author{Marcin Mierzejewski}
\affiliation{Department of Theoretical Physics, Faculty of Fundamental Problems of Technology, Wroc\l aw University of Science and Technology, 50-370 Wroc\l aw, Poland}
\author{Janez Bon\v ca}
\affiliation{Department of Physics, Faculty of Mathematics and Physics, University of Ljubljana, SI-1000 Ljubljana, Slovenia}
\affiliation{Department of Theoretical Physics, J. Stefan Institute, SI-1000 Ljubljana, Slovenia}

\begin{abstract}
We study sample-to-sample fluctuations of the gap ratio in the energy spectra in finite disordered spin chains. The chains are described by the random-field Ising model and the Heisenberg model. We show that away from the ergodic/nonergodic crossover, the fluctuations are correctly captured by the Rosenzweig-Porter (RP) model. However, fluctuations in the microscopic models significantly exceed those in the RP model in the vicinity of the crossover. We show that upon introducing an extension to the RP model, one correctly reproduces the fluctuations in all regimes, i.e., in the ergodic and nonergodic regimes as well as at the crossover between them. Finally, we demonstrate how to reduce the sample-to-sample fluctuations  in both studied microscopic models.
\end{abstract}
\maketitle

\section{Introduction} 

Switching on interactions in  low--dimensional Anderson insulators leads  through the interplay between  the quantum interference and many--body interactions to fascinating new phenomena. While turning  on interactions at small disorder delocalizes the system, the strong disorder is believed to cause  many--body (MBL) localization~\cite{gornyi05,basko_aleiner_06,oganesyan_huse_07,AbaninRMP}  at least on finite--size lattices.  Even though    research in this field  predominantly  focused on a few simplest prototype model Hamiltonians for MBL, such as the disordered  XXZ model~\cite{barisic_prelovsek_10,luitz15,luitz16,torres15,sirker14,pal10,bera15,Hauschild_2016,Devakul2015,bertrand_garciagarcia_16,Husex2017,Doggen2018}, the type of the transition and even the existence  of the MBL phase in the thermodynamic limit are  still under intense consideration~\cite{suntajs2020e,suntajs2020, sirker_2020, Luitz20, Vasseur20,Polkovnikov21a,sieran2021a,vidmar2021,Polkovnikov21,Huse21,sels2021,Sirker21}.  The latest results based on the studies of the avalanche instability suggest that the transition to the MBL phase  occurs for much stronger disorder then it follows from the previous numerical studies of  finite systems \cite{Huse21}.

At small disorder, the quantum many-body system is ergodic; its energy spectrum can be analyzed within the framework of the random matrix theory (RMT), while the eigenstate thermalization hypothesis can describe 
the relaxation of physical observables of a closed system~\cite{deutsch_91, srednicki_94, rigol_dunjko_08, dalessio_kafri_16, mori_ikeda_18, deutsch_18,dalessio_kafri_16, santos2010, Beugeling2014, Steinigeweg2014, Kim_strong2014, mondaini_rigol_17, jansen_stolpp_19, leblond_mallayya_19, mierzejewski_vidmar_20, brenes_leblond_20, richter_dymarsky_20, schoenle_jansen_21, brenes_pappalardi_21}. Increasing the disorder strength leads to the ergodicity breakdown that is reflected in the departure from the RMT
prediction and can be studied  through the behavior of  different ergodicity indicators such as the anomalous level statistics and the eigenstate entanglement entropies~\cite{suntajs2020e,suntajs2020}, the fidelity susceptibility~\cite{Polkovnikov21a}, the anomalous  distribution of observable matrix elements~\cite{Panda2020, corps_molina_21}, the opening of the Schmidt gap~\cite{gray_bose_18},  the gap in the spectrum of the eigenstate one-body density matrix~\cite{bera15}, and the correlation-hole time in the survival probability reaching the Heisenberg  time $t_{\rm H}$~\cite{schiulaz_torresherrera_19}.

Strong disorder is also reflected in  unusual transport properties of the system, such as the onset of slow relaxation \cite{znidaric08,bardarson12,kjall14,serbyn15,luitz16,serbyn13_1,bera15,altman15,agarwal15,gopal15,znidaric16,mierzejewski2016,lev14,lev15,barisic16,bonca17,bordia2017_1,zakrzewski16,protopopov2018,sankar2018,zakrzewski2018}, subdiffusive transport~\cite{lev15, agarwal15, luitz16, khait_gazit_16, znidaric16, luitz_barlev_17, bera_detomasi_17,kozarzewski18,prelovsek217,prelovsek2018a} and an approximate $1/\omega$ scaling of the spin density spectral function~\cite{mierzejewski2016, serbyn2017, Polkovnikov21a}.
Recently, the latter scaling was analyzed  in a framework of a phenomenological theory based on the proximity to the local integrals of motion of the Anderson insulator, which describes  the dynamics of the observables at infinite temperature~\cite{vidmar2021}. 

The shift  from the ergodic regime by increasing disorder is accompanied by the increase of fluctuations of various physical  quantities.  Non--Gaussian fluctuations of the  resistivity $\rho$ follow  the departure from the ergodic regime  while   the probability distribution of $\rho$ reveals fat tails that appear generic for different disordered many--body models \cite{mierzejewski2020}. The time evolution of fluctuations of the return probability has been recently used to 
explore the possibility of the existence of different types of MBL phases~\cite{Bera2019}. Lately, the level statistics of disordered interacting quantum system has been analysed throughout  the crossover
 from the ergodic to many--body localized phase~\cite{zakrzewski2019,zakrzewski2020}.

In this work, we discuss sample-to-sample fluctuations of the  gap ratio \cite{zakrzewski2020} in two basic models used for studying crossover from the ergodic system with  the GOE level statistics to the non-ergodic one with the Poisson statistics  in disordered finite spin chains: the random-field Ising model and the
random-field Heisenberg chain.  In the vicinity of the crossover,  both models display similar distribution of fluctuations. We show that this distribution can be adequately  described  within the framework of the  Rosenzweig-Porter (RP) \cite{RP1960,Serna2019}  model with a modification that accounts for different realizations of disorder in finite systems. Further on, we refer to the modified model as the  Rosenzweig-Porter model with sample fluctuations (RPSF model). The only difference between both models consists in that the variance of the off-diagonal matrix elements in RPSF model is not set directly by the dimension of the Hilbert space (as it is the case in the RP model)  but is drawn from the normal distribution with its  spread given  by the dimension of the Hilbert space. We further show that the standard RP and the RPSF models are equivalent in the thermodynamic limit. Using this property, we show that the sample-to-sample fluctuations may be reduced also in both spin chains, provided one studies the gap ratio not for a fixed disorder strength but for fixed norms of the diagonal and off-diagonal parts of the Hamiltonian.

\section{Models and fluctuations of the gap ratio } 

We discuss the level statistics in two models which have been commonly studied in the context of the many-body-localization. First, we consider the random-field Ising chain
\begin{equation}
H_I = \sum_{i} \left(J+\delta J_i\right)S_i^zS_{i+1}^z + \sum_i h_i S_i^z+f\sum_i S_i^x, \label{hamisI}
\end{equation}
 where we set the transverse field $f/J=0.5$. In order to break the integrability of the system without disorder ($W=0$), we add also a small randomness to the coupling constant, $\delta J_i\in\left[ -W_J,W_J\right]$ with $W_J/ J=0.2$.  The second model is the random-field Heisenberg chain, which has been mostly used in the numerical calculations
\begin{equation}
H_H = \sum_{i}  J \vec{S}_i \cdot \vec{S} _{i+1} + \sum_i h_i S_i^z. \label{hamis}
\end{equation}
In both cases we consider chains with $L$ sites and with periodic boundary conditions.  We also set $J=1$ as the energy unit and assume uniformly distributed random field $h_i\in\left[-W,W\right]$. 

In order to distinguish between ergodic and localized phases, we  follow a commonly used procedure and  calculate the gap ratio (introduced in Ref.~\cite{oganesyan_huse_07}), defined as $r_n=\min\{\delta_n,\delta_{n-1}\}/\max\{\delta_n,\delta_{n-1}\}$, where $\delta_n=E_{n+1}-E_n $ is the gap between the consecutive energy levels. Usually, one investigates $r_n$ that is  
averaged over multiple energy levels from the middle of the spectrum as well as over various realizations of disorder, i.e. over multiple sets $\{h_1,...,h_L \}$. One expects
 $r_{Poisson} \approx 0.386$ for the localized phase and  $r_{GOE} \approx 0.5295$ for the ergodic one.
 
 In this work, we focus on the gap ratio that is averaged only over 
 the energy levels with a given realization of disorder and discuss fluctuations of such quantity between various realizations of disorder (sample-to-sample fluctuations).
 This problem has previously been studied for the Heisenberg chain in Ref. \cite{zakrzewski2020} which reported large fluctuations  of the gap ratio in the vicinity of the crossover between  GOE and  Poisson level statistics. Such fluctuations hinder accurate finite-size scaling and  precise location of the crossover.  
  As a main result of this work, we show that the sample-to-sample fluctuations may be reduced via appropriate identification of  {\it more ergodic} and  {\it less ergodic} samples.
  In the thermodynamic limit, the latter feature is expected to be uniquely determined by $W$, what is however not the case for finite systems. 
  
  To this end, we study  the  gap ratio $r_S$ which is a mean value of $r_n$ studied separately for each disorder realization \mbox{$r_S=\frac{1}{N_r} \sum_{i=1}^{N_r} r_n$}, where $N_r+2$ is the number of energy levels taken from the middle part of the spectrum. If not otherwise stated, we use $N_r=\frac{1}{3}Z$ where $Z$ is 
  the dimension of the Hilbert space. 
\begin{figure}[!htb]
\includegraphics[width=0.9\columnwidth]{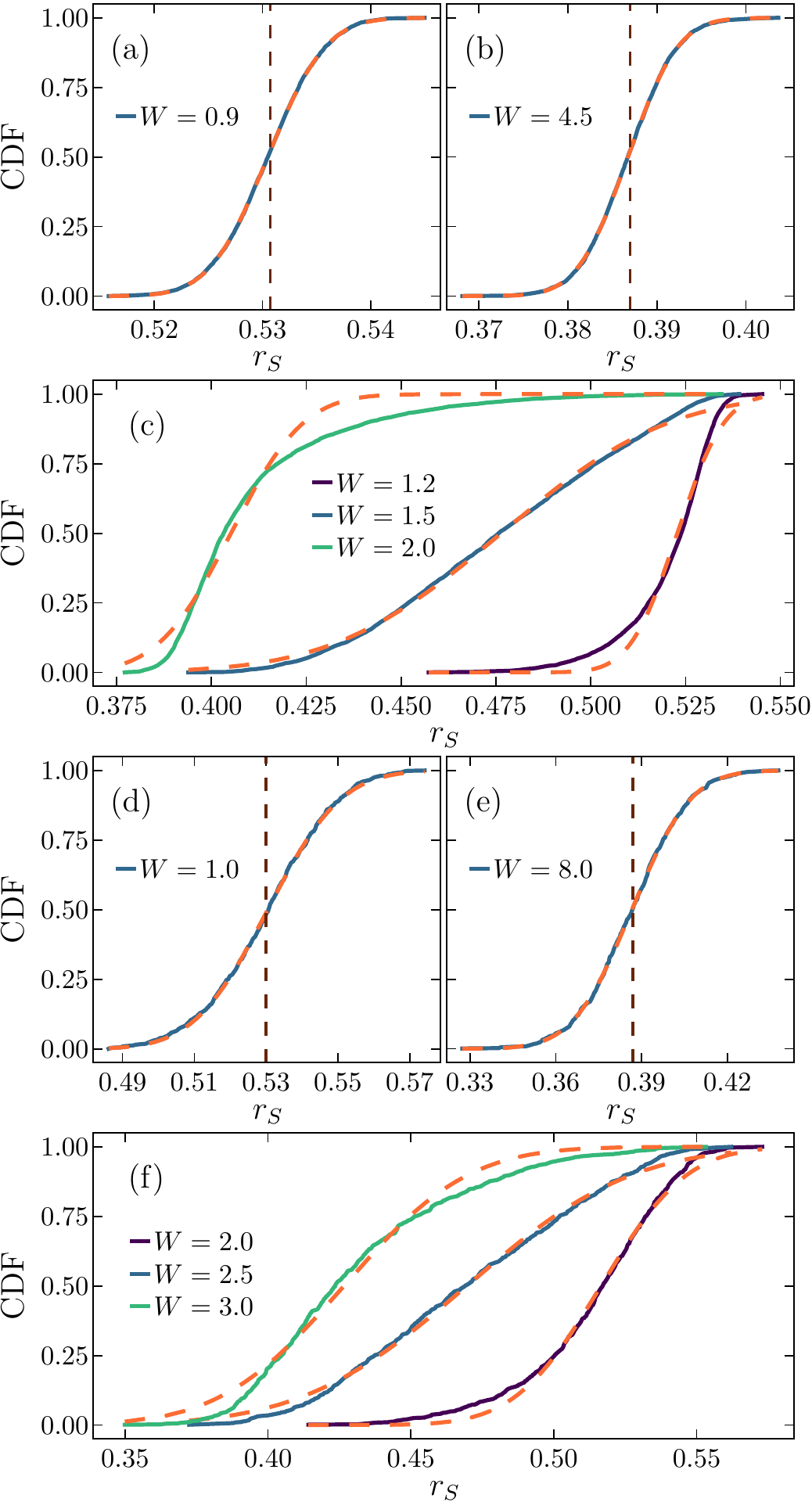}

\caption{Cumulative distribution functions (CDF) of $r_S$ obtained for different realizations of disorder and fixed $W$. 
CDF's for various $W$ are fitted with the error function (dashed curves). Results in  (a),(b),(c) are for the Ising model with  $L=14$ and $4000$ disorder realizations; (d),(e),(f) show results for the Heisenberg chains with $L=16$ and 700 samples of disorder.}
\label{fig1}
\end{figure}

\section{Fluctuations of the gap  ratio}

Numerical studies of the Heisenberg chain have been carried out in the sector with the total spin projection $S^z_{tot}=0$. Since $S^z_{tot}$ does not commute with the Ising Hamiltonian,
$H_I$, we use the full Hilbert space in the latter case. Consequently, in the case of the Heisenberg chain, we may access larger system sizes than for the Ising chain.
Figure \ref{fig1} shows cumulative distribution functions of $r_S$ obtained from multiple realizations of disorder for fixed $W$ and various panels
correspond to different values of $W$. In both models, results for weak  disorder [Figs. \ref{fig1}(a) and \ref{fig1}(d)] or strong disorder  [Figs. \ref{fig1}(b) and \ref{fig1}(e)]  can be accurately fitted by the error function,  and the fits are shown as  dashed curves. In both regimes one observes  Gaussian sample-to-sample fluctuations centered at $r_{GOE} $ (for weak disorder) or $r_{Poisson}$ (for strong disorder).
 However, for intermediate disorders  [Figs. \ref{fig1}(c) and \ref{fig1}(f)] the distributions visibly deviate from the error functions. In particular, the values of $r_S$ obtained for the Ising model at $W=1.5$  
span almost the whole window between $r_{Poisson}$  and  $r_{GOE} $, see  Figs. \ref{fig1}(c). Similar observation holds true also for the Heisenberg model  at  $W=2.5$, as it is shown in Fig.  \ref{fig1}(f).

\begin{figure}[!htb]
\includegraphics[width=0.9\columnwidth]{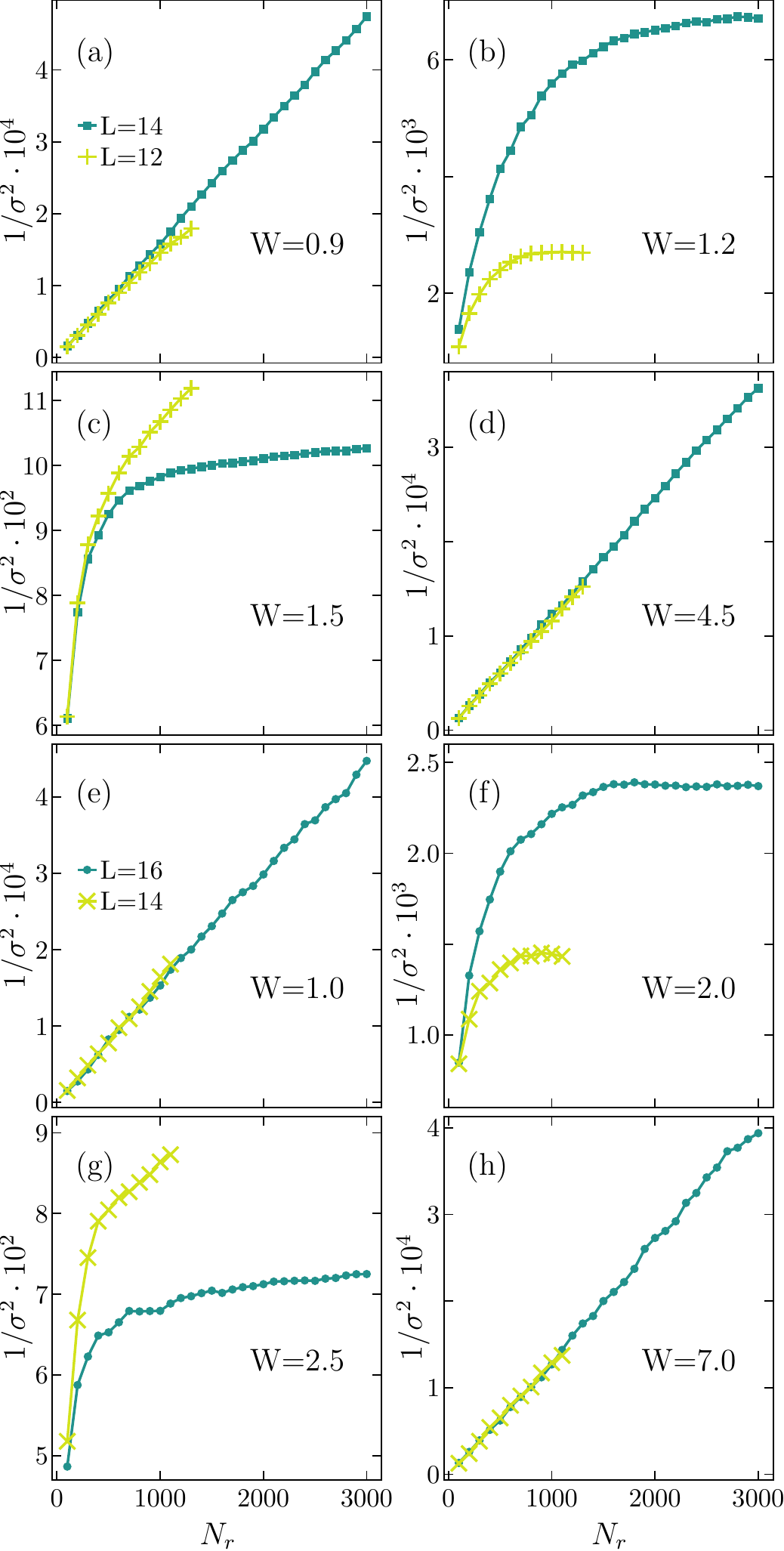}
\caption{The inverse variances of distributions shown in Fig. \ref{fig1} vs. number of energy levels, $N_r$, used for averaging of $r_n$. Results in (a) – (d)  are for the Ising model with  $L = 12$ or $L=14$ sites and $N_d=4000$ realizations of disorder. Plots (e) – (h) show the same but for Heisenberg model with $L = 14$ or $L=16$ sites and $N_d=700$ realizations of disorder.}
\label{fig2}
\end{figure}

In order to identify the origin of these fluctuations, we calculate the variances of the distributions shown in Fig. \ref{fig1},
$\sigma^2=\langle r^2_S \rangle_d - \langle r_S \rangle^2_d $, where $\langle ... \rangle_d$ means averaging over $N_d$ samples of disorder. 
 In Fig.~\ref{fig2} we show how this quantity depends on the number of  energy levels,  $N_r$, used for evaluation of $r_S$ for each sample. For weak or strong disorder, the variances 
 scale as  $\sigma^2 \propto 1/N_r$  while the dependence of   $\sigma^2$ on $L$ is rather insignificant. This behavior is shown in Figs. \ref{fig2}(a) and \ref{fig2}(d) for the Ising model and in Figs. \ref{fig2}(e) and \ref{fig2}(h) for the Heisenberg chain.  Therefore, in the regime of weak or strong disorder  the fluctuations of the gap ratio seem to have  purely statistical origin which can be linked solely to the number 
 of energy levels which are accessible in finite systems.   The correlations between
 $r_n$ and $r_m$ should not be essential for sufficiently distant $m$ and $n$. Then,  $r_S=1/N_r \sum_{i=1}^{N_r} r_n$ is a random variable. Its distribution tends toward the normal distribution for large $N_r$ with the variance that is inversely proportional to $N_r$.

Such picture breaks down for the intermediate disorder strengths, as it can be observed from Figs. \ref{fig2}(b), \ref{fig2}(c) for the Ising model and  Figs. \ref{fig2}(f), \ref{fig2}(g) for
the Heisenberg chain.  The  distributions are much broader than for the previously discussed cases, $1/\sigma^2$ almost saturates for large $N_r$ and shows strong dependence on the system size $L$.
 Departure of  the distribution of $r_S$ from the normal distribution for large $N_r$  suggests that certain realizations of disorder $\{h_1,...,h_L\} $ are more
 relevant for the "ergodic system" while the others are more relevant for the "localized system" despite being drawn for the same disorder strength $W$.  In other words, we expect that for the accessible system sizes, the sequence $\{h_1,...,h_L\} $ is too short to be fully specified by a single quantity, $W$.  
 
In order to test this conjecture and to decrease the sample-to-sample fluctuations of $r_S$ we introduce a new parameter to describe the properties of the random sequence  $\{h_1,...,h_L\} $,
\begin{equation}
\mathcal{V}=Z \delta_{nd}^2/\delta_{d}^2
\label{bignu}
\end{equation}
where $\delta_{d}$ and  $\delta_{nd}$ are, respectively, the variances of the diagonal and off-diagonal elements of the Hamiltonian matrix,  $H_{\bar{s},\bar{s}'}=\langle \bar{s}|H |\bar{s'} \rangle$, in the real-space basis 
\mbox{$|\bar{s} \rangle=|S^z_1,S^z_2,...,S^z_L\rangle$. }
The ratio is rescaled by the dimension of the Hilbert space, $Z$,  in order to have a non-zero value for $L \to \infty$.  

The result for $\mathcal{V}$  can be obtained from the high-temperature expansion. 
In the case of the Heisenberg model, the variance of the off-diagonal part equals, up to the $1/Z$ factor,  the Hilbert-Schmidt norm of the spin-flip term,  $ \delta^2_{nd}=L/(8Z)$.
The diagonal variance is determined by the norms of the $S^z S^z$-term and the random field term,  $ \delta^2_{d}=L/16+1/4 \sum_{i=1}^L h^2_i$.  Since $h_i$ are independent random variables, one finds 
\begin{equation}
 \delta^2_{d}= L\left( \frac{1}{16}  +\frac{W^2}{12}+ \frac{1}{4\sqrt{L}} y \right), \label{new0}
\end{equation}
where $y$ is  a random variable which for large $L$ is described by a normal distribution with mean zero and the variance  $4 W^4/45$. Then, one may calculate $\mathcal{V}$ for 
the Heisenberg model in the thermodynamic limit
\begin{equation}
\mathcal{V}^H_\infty=\frac{6}{3+4W^2}.
\label{V_inf_H}
\end{equation}
Similar analysis applied  for the Ising chain yields 
\begin{equation}
\mathcal{V}^{I}_{\infty}=\frac{12f^2}{3+W_J^2+4W^2}.
\label{V_inf_TFI}
\end{equation}
The properties of a finite system are not determined by $W$ but rather by the sequence of random potentials $\{h_1,...,h_L\} $. Then, the question is whether the physical
properties of a finite system with a fixed sequence are more accurately encoded in the value of $W$ (used to generate the sequence) or in the
ratio $\mathcal{V}$. In order to answer this question one needs to check whether the sample-to-sample fluctuations are smaller within results obtained for fixed $W$ or for fixed $\mathcal{V}$. Further on we demonstrate that the latter case holds true.
The motivation for discussing $\mathcal{V}$  comes from that it is directly related to the norm of the perturbation term relatively to the norms of the other terms of the Hamiltonian. In other words, it allows to distinguish "more localized" samples with larger $ \sum_{i=1}^L h^2_i$  from "more ergodic" cases where the latter quantity is smaller. An essential observation following from Eq. (\ref{new0})  is that in the thermodynamic limit there is one-to-one correspondence between
the values of $W$ and $\mathcal{V}$, so the system's properties are fully specified by either of these quantities.
An additional support for using $\mathcal{V}$ is discussed in the subsequent section, where we introduce a phenomenological model which accurately reproduces broad, non-Gaussian fluctuations of $r_S$ in finite systems.

\section{Phenomenological model for the sample-to-sample fluctuation of  gap ratio. }
 Our phenomenological approach is motivated by the Rosenzweig-Porter model \cite{RP1960,Serna2019}, where the ergodic-nonergodic crossover originates from different variances of the diagonal and off-diagonal  elements of random-matrices.
  In that model, one considers random matrices, $H_{\bar{s},\bar{s}'}$,   such that  diagonal elements and all off-diagonal  elements  in the upper triangle of the symmetric matrix  are  sampled from a normal (Gaussian)  distribution with  $\delta_{d}^2=1$ and $\delta_{nd}^2= Z^{-\gamma}/2$, respectively,  and where the crossover takes place at $\gamma=2$. In order to account for the finite-size fluctuations discussed in the preceding section, we consider a generalisation to the RP model with sample fluctuations (RPSF),  where  $\delta_{nd}$ is not a constant for different samples  but rather a random variable that changes from sample to sample.
% In that model, one considers random matrices, $H_{\bar{s},\bar{s}'}$,   such that $\delta_{d}^2=1$ and $\delta_{nd}^2= Z^{-\gamma}/2$ and the crossover takes place
%at $\gamma=2$. In order to account for the finite-size fluctuations discussed in the preceding section, we consider   $\delta_{nd}$ that is not a constant but rather a random variable that changes from sample to sample. 
 We assume that   $\delta_{nd}$ is drawn from the normal distribution with the variance $Z^{-\gamma}/2$ with the probability density function 

 \begin{equation}
 f_{\delta}(\delta_{nd})=\frac{2}{\sqrt{\pi Z^{-\gamma}}} \exp \left( -\frac{\delta^2_{nd}}{Z^{-\gamma}} \right), \quad \quad \delta_{nd} \ge 0.
 \label{new1}
 \end{equation}
 
Note that despite this generalisation, RPSF model remains a single parameter ($\gamma$)  model,  just like the standard RP model. 
Within the RP and RPSF models one generates dense matrices where all matrix elements are nonzero. However, in microscopic models discussed in the preceding section, the number of nonzero off-diagonal elements grows as $\sim Z log(Z)$.

\begin{figure}[!htb]
\includegraphics[width=0.9\columnwidth]{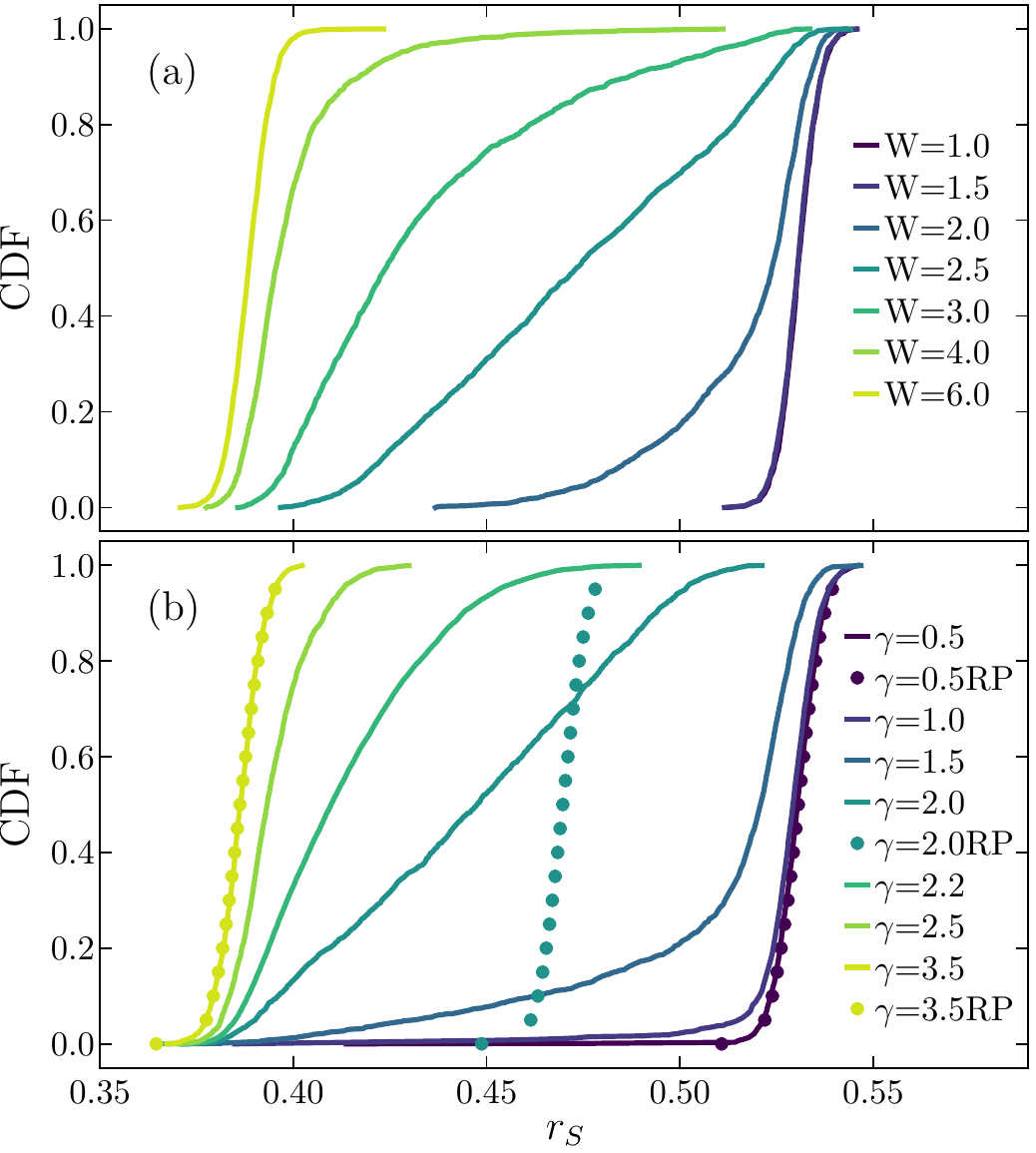}
\caption{ Cumulative distribution functions for $r_S$. Panel (a) shows numerical results for the random-field Heisenberg chain with $L=16$ sites and various $W$. Points in (b) show CDF's for the standard  Rosenzweig-Porter model and lines show results for the latter model that is generalized to account for the sample-to-sample fluctuations, the RPSF model. In order to obtain results in (b) we have generated random $3500 \times 3500$ matrices. 
}
\label{fig3}
\end{figure}

  In order to carry out simulations of such model  first we draw random  $\delta_{nd}$ and then for fixed $\delta_{nd}$ and 
  $\delta_{d}=1$ we generate normally distributed real-valued matrix elements. 
In figure \ref{fig3} we compare the distributions of $r_S$ obtained for the random-field Heisenberg chain with $L=16$ sites [panel (a)] and for the RP or RPSF models [panel (b)]. In the latter case, we have diagonalized  random $3500 \times 3500$ matrices generated according to the procedure discussed in the preceding paragraph. We recall that in the case of the Heisenberg chain, the dimension of the Hilbert space is approximately 12000. One observes that the standard RP model [points in Fig. \ref{fig3}(b)] correctly captures the distributions of $r_S$ only for parameters which are  far away from the GOE/Poisson crossover, i.e., for $\gamma \ll 2$ or $\gamma \gg 2$. Even though random matrices are smaller than the matrices representing the Heisenberg Hamiltonian, the standard RP model fails to reproduce the broad distributions observed in the vicinity of the crossover at $\gamma =2$. In contrast, the generalized RPSF model correctly reproduces the numerical studies of the microscopic model for arbitrary $\gamma$. Interestingly, both versions of the RP model give almost indistinguishable results for $\gamma \ll 2$ or $\gamma \gg 2$. Therefore, the generalization affects predominantly  the sample-to-sample fluctuation observed in the vicinity of the crossover.
 
 Next we demonstrate that the sample-to-sample fluctuation in RPSF vanish in the termodynamic limit, $Z \to \infty$, and  discuss how these fluctuation depend on the system size. To this end we transform the random variable 
 $\delta_{nd}$ in Eq. (\ref{new1}) and introduce $\Gamma$ such that $Z^{-\Gamma}/2=\delta^2_{nd}$. It is clear that $\Gamma$ plays that same role in  RPSF model as the exponent $\gamma$ in the standard RP model except that $\Gamma$ changes from sample to sample. In order to obtain the probability density function $ f_{\Gamma}(\Gamma)$ we compare the cumulative distribution functions
 \begin{equation}
 \int_{\Gamma}^{\infty} {\rm d \Gamma'}  f_{\Gamma}(\Gamma')=\int_{0}^{X(\Gamma)}  {\rm d \delta_{nd}} \;\;  f_{\delta}(\delta_{nd}), 
  \label{new2}
 \end{equation}
where $X(\Gamma)=Z^{-\gamma/2}/\sqrt{2}$. Differentiating Eq. (\ref{new2}) with respect to $\Gamma$ one obtains
\begin{eqnarray}
f_{\Gamma}(\Gamma)&=&\ln(Z) g[\ln(Z)(\Gamma-\gamma)], \nonumber \\
g(x)&=&\frac{1}{\sqrt{2 \pi}} \exp \left( -\frac{x}{2}  -\frac{1}{2} e^{-x}  \right).
\end{eqnarray} 
Note that  $g(x)$ is normalized and the corresponding  cumulative distribution gives $G(x) = 1-\mathrm{erf}(e^{-x/2}/\sqrt{2})$ where erf is the Gaussian error function.   The distribution function  $f_{\Gamma}(\Gamma)$ approaches the delta function $\delta(\Gamma-\gamma)$ for $Z \to \infty$. Consequently, the standard RP 
and the RPSF models are equivalent in the thermodynamic limit.  One may also find the finite size fluctuations
\begin{equation}
\langle \Gamma^2 \rangle-\langle \Gamma \rangle^2 =\frac{\pi^2}{2 \ln(Z)^2}.
\end{equation} 
Interestingly, the standard deviation of $\Gamma$ decays as the inverse of the linear system's dimension, $\ln(Z)$, even though such dimension does not explicitly enter the RPSF model. To conclude this section, we note that
a simple extension to the RP model allows reproducing the sample-to-sample fluctuations in a finite system, whereas, in the thermodynamic limit, both models (RP and RPSF) are equivalent. 
 
\section{Numerical results for disordered Ising and Heisenberg chains}

\begin{figure}[!htb]
\includegraphics[width=0.9\columnwidth]{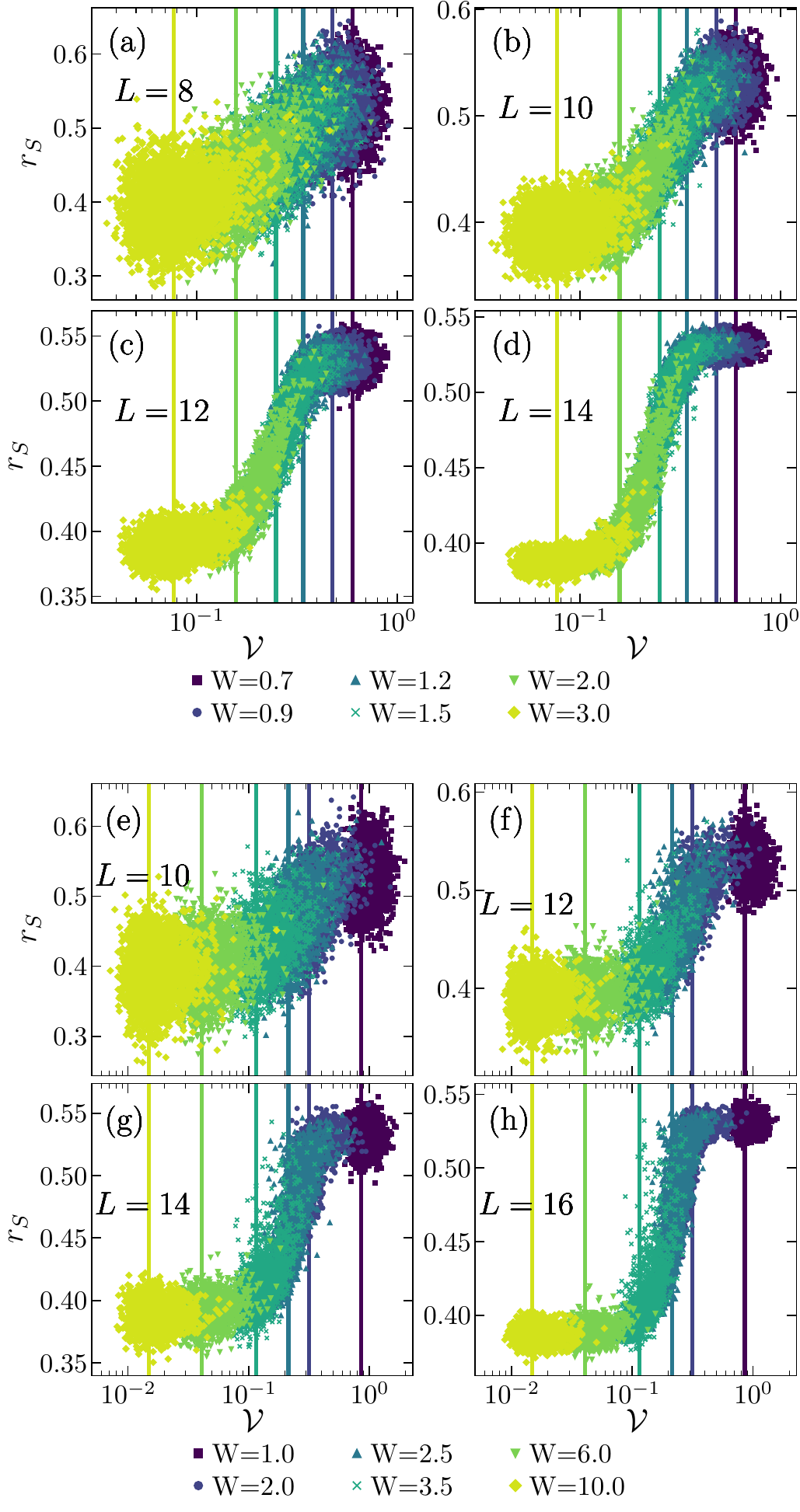}
\caption{Correlations between the gap ratio $r_S$ and $\mathcal{V}$, see Eq. (\ref{bignu}). Each point corresponds to a single realization of disorder and different symbols distinguish between values of $W$ for
which the sequence of random fields, $h_1,...,h_L$, was drawn.  Panels (a)-(d) and (e)-(h) show results for the Ising and Heisenberg models, respectively.
The vertical lines mark results the values of $\mathcal{V}$ in the thermodynamic limit,  given  by Eqs. (\ref{V_inf_TFI}) and  (\ref{V_inf_H}), respectively. }
\label{fig4}
\end{figure}

\begin{figure}[!htb]
\includegraphics[width=0.9\columnwidth]{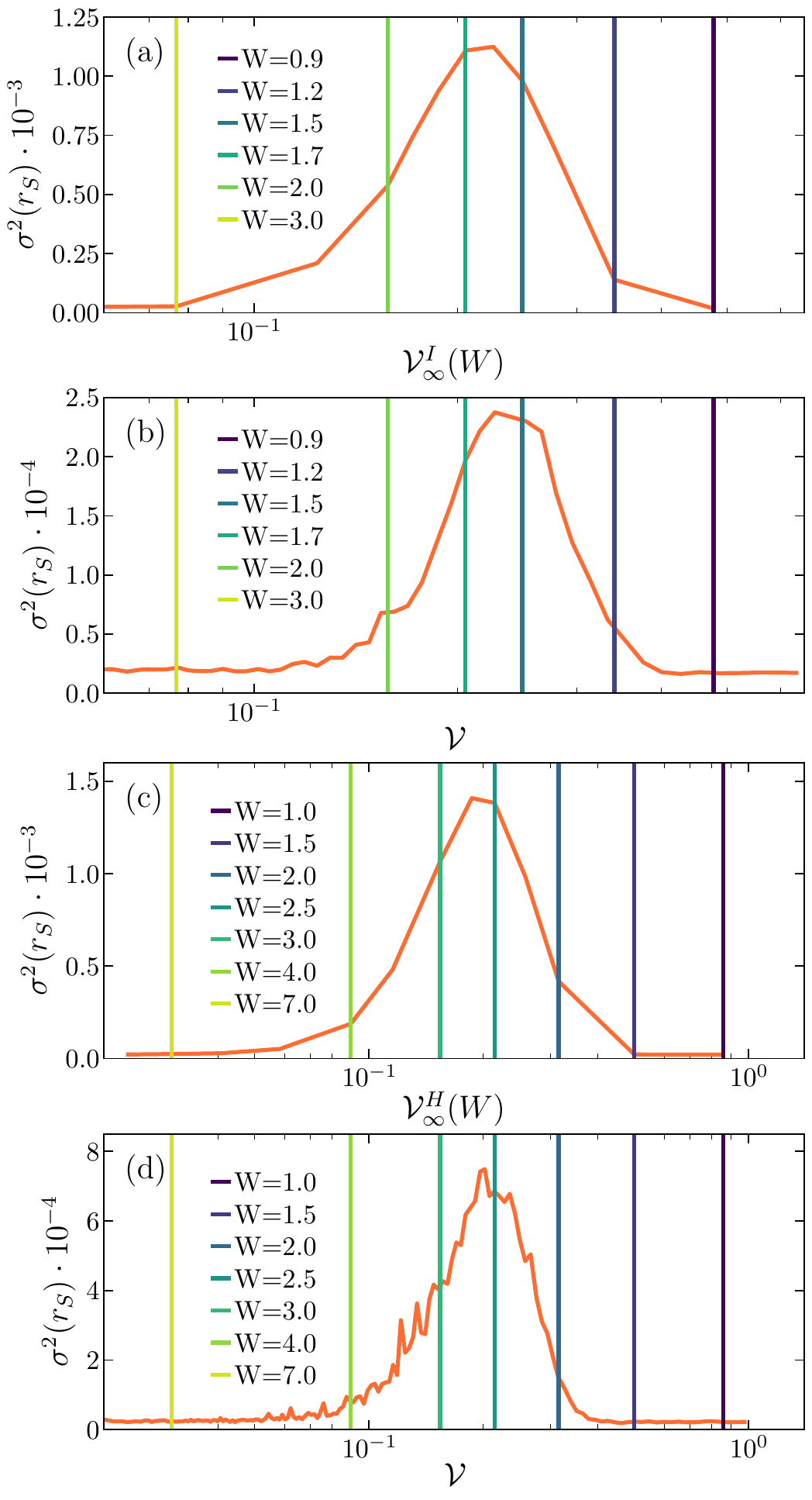}
\caption{The variance of the gap ratio obtained for fixed $W$ in (a) and (c) or fixed $\mathcal{V}$ in (b) and (d). Results in (a), (b) are for the Ising chain and in (c),(d) for the Heisenberg model. 
For the clarity, results in (a) and (c) are plotted as functions of   $\mathcal{V}^I_{\infty}(W)$ and $\mathcal{V}^H_{\infty}(W)$.
The vertical guidelines mark the values of $\mathcal{V}$ in the thermodynamic limit for selected $W$ listed in the legend. 
}
\label{fig5}
\end{figure}

Here we demonstrate that using  results from the proceeding section one may reduce sample-to-sample fluctuations also in both microscopic models. In Fig.~\ref{fig4} we show correlations between the  gap ratio, $r_s$, and $\mathcal{V}$, see Eq. (\ref{bignu}), where both quantities  are calculated separately for each disorder realization. We use different symbols to distinguish between disorders drawn for various
$W$ and vertical guidelines mark $\mathcal{V}^I_{\infty}$ [Eq. (\ref{V_inf_TFI})] and $\mathcal{V}^H_{\infty}$  [Eq. (\ref{V_inf_H})] obtained for the Ising and the Heisenberg chains, respectively. One observers that results
obtained for various realizations of disorder form a single sigmoid-like curve. The points are scattered in both directions. The vertical scattering means that $\mathcal{V}$ does not fully specify $r_S$ in finite systems, whereas the horizontal scattering means that $W$ does not fully specify $\mathcal{V}$. The scattering in the horizontal direction is responsible for  fluctuation of 
$r_s$ which are intensified on the steep section of the sigmoid, i.e, in the vicinity of the crossover between the Poisson and the GOE level statistics. 

The latter  contribution to $\sigma$ may be eliminated once one studies fluctuations for fixed  $\mathcal{V}$ instead of fixed $W$. To this end, in each step of simulations, we first randomly choose $W$,
$0.7<W<3.5$ for the Ising chain and $1<W<8$ for the Heisenberg model,  we draw the random fields $h_i$ and then construct the Hamiltonian and obtain $\mathcal{V}$ and $r_S$. The results are identical 
to those in Fig.~\ref{fig4} except that now points almost uniformly cover the entire  range of  $\mathcal{V}$. Then in narrow windows of $\mathcal{V}$ we calculate the variances, $\sigma^2$.
Results are shown in figures \ref{fig5}(b) and  \ref{fig5}(d) for the Ising and Heisenberg chains, respectively.  In panels (a) and (c) we show the spreads of distributions in Fig. \ref{fig1}, i.e., distributions 
generated in a standard way for fixed $W$. In order to facilitate the  comparison of both types of results in  (a) and (c), we plot $\sigma^2$ vs.   $\mathcal{V}^I_{\infty}(W)$ and $\mathcal{V}^H_{\infty}(W)$, respectively.
For the Ising model, we observe that  $\sigma^2$ obtained from fixed  $\mathcal{V}$ is approximately five times smaller than  $\sigma^2$ obtained from fixed  $W$. In the  case of the Heisenberg model, the reduction
of fluctuations is less significant but still clearly visible.

Following the suggestion in Ref.~\cite{Huse22}, we have concentrated our efforts on statistics of energy levels in the crossover regime from GOE to MBL  observed on finite--size lattices. There is particular attention devoted to the intermediate crossover regime as it appears on finite lattices in the hope that a better understanding of this regime would yield additional new information about the nature of the MBL phase transition\cite{Crowley_2020,Huse22} in the thermodynamic limit. 
Our research was also inspired by the notion that in the regime of strong disorder where the MBL phase is predicted to occur, the finite--size effects critically affect results obtained from exact diagonalization approaches on a small lattice sizes\cite{Sierant2020,Panda2020,Vasseur2021}. Our approach demonstrates how to reduce the finite-size sample-to-sample fluctuations and it does not relay on any particular  picture that holds in the thermodynamic limit.

\section{Conclusions}

In finite disordered systems, we have studied the sample-to-sample fluctuations of the gap-ratio, $r_S$, which are expected to show the GOE/Poisson crossover. In order to single-out generic model-independent properties, we have numerically studied the random-field Heisenberg chains as well as the random-field Ising model. Far from the crossover, the fluctuations of $r_S$ are set by the dimension of the Hilbert space. Consequently, the standard RP model correctly reproduces the fluctuations obtained for microscopic systems. Random matrices of comparable sizes as the matrices of the microscopic Hamiltonians give rise to similar fluctuations. However, this is not the case in the vicinity of the crossover when the fluctuations in the microscopic models significantly exceed the results for the standard RP model. This discrepancy is because the relatively small sets of random fields, $h_1,..., h_L$ may be very different despite being drawn for the same disorder strength, $W$. As a result, specific samples drawn for finite system appear {\it more ergodic} while the others are {\it more localized}. We have demonstrated that this feature may be implemented in the generalized RP model rather straightforwardly. In its generalized version, i.e, in the RPSF model, the variance of the off-diagonal matrix elements is a normally-distributed random variable that changes from sample to sample. The width of the latter distribution is the same as the constant variance in the standard RP model. The generalized RPSF model accurately captures the distributions of the gap ratio in finite  chains essentially for all studied regimes. Moreover, far away from 
the GOE/Poisson crossovers, both versions of the Rosenzweig-Porter model give almost identical distributions.

Utilizing the latter result, we have demonstrated that the sample-to-sample fluctuations in the finite microscopic model can be reduced. Instead of studying the distributions of the gap ratio for fixed $W$, we have evaluated  them in terms of  a fixed ratio of the Hilbert Schmidt norms of the diagonal and off-diagonal parts of the Hamiltonian. The norm of the random-field term depends on $W$ but also 
accounts for differences between various realizations  of the random fields. The latter ratio of norms is a counterpart of the ratio of variances in the RPSF model. We have shown that this procedure significantly reduces sample-to-sample fluctuations in the Ising model, while the reduction for the Heisenberg model is less pronounced.

\acknowledgments

We acknowledge the support by the National Science Centre, Poland via project 2020/37/B/ST3/00020 (B.K and M.M.), the support by the Slovenian Research Agency (ARRS), Research Core Fundings Grant P1-0044 (J.B.), and the support from  the Center for Integrated Nanotechnologies, a U.S. Department of Energy, Office of Basic Energy Sciences user facility (J.B.). The numerical calculations were partly carried out at the facilities of the Wroclaw Centre for Networking and Supercomputing.

%----------------------------------------------------------------------------------------
\bibliography{references_ergtransition,references}
\end{document}